# A Rapid Compression Study of the Butanol Isomers at Elevated Pressure

*B. Weber and C.J. Sung*

*Department of Mechanical Engineering*
*University of Connecticut, Storrs, CT 06269, USA*

Investigation of the autoignition delay of the butanol isomers has been performed at elevated pressure of 15 bar and low to intermediate temperatures of 725–870 K. Stoichiometric mixtures made in nitrogen/oxygen air were studied. For the temperature and pressure conditions in this study, no NTC or two-stage ignition behavior were observed. The reactivity of the isomers of butanol, in terms of inverse ignition delay, was ranked as $n$-butanol $>$ $sec$-butanol $\sim$ $iso$-butanol $>$ $tert$-butanol. Predictions of the ignition delay by several kinetic mechanisms available in the literature generally over-predict the ignition delays.

## 1. Introduction

Recent environmental and geo-political concerns have led to a renewed push to develop alternative sources for fuels. In particular, many efforts have been made to reduce the consumption of petroleum-based fuels in vehicles. Automobile manufacturers have improved the volumetric fuel economy of their fleets significantly over the last decade, while also enabling the use of alternative fuels such as ethanol. Unfortunately, ethanol is less than an ideal replacement for gasoline in current engines, due to its lower volumetric energy density, propensity to absorb water, and feedstocks which may consume world food supply [1,2].

Therefore, a second generation of alternative fuels is being developed to help alleviate the concerns with using ethanol. One of the most promising of these fuels is *n*-butanol. *n*-Butanol has much closer energy density to that of gasoline, making it more suitable as a blending component or drop-in replacement for gasoline. It is less hygroscopic than ethanol, and technologies are being developed to produce *n*-butanol from many feedstocks, including crops that can be grown on marginal land not suited for food crops [2].

In addition, there are 4 isomers with the chemical formula $C_4H_9OH$ – *n*-, *sec*-, *tert*-, and *iso*-butanol. The butanol system comprises the smallest alcohol system with primary, secondary, and tertiary type alcohol groups. Moreover, the $C_4$ chain would be able to display intramolecular isomerization chemistry that is important in larger fuels. This makes the butanols a good test case to develop models for higher alcohols.

The number of studies of *n*-butanol has increased dramatically in the last year. A small sampling of recent results includes flame speeds [3], ignition delays [4,5] and pyrolysis studies [6]. Although the sheer number of studies of the isomers of *n*-butanol is significantly less, similar types of results are available [7-10]. However, there is a scarcity of data at higher pressures and lower temperatures, especially for ignition delays. In this study, autoignition delay results collected using a heated Rapid Compression Machine (RCM) are presented for the four isomers of butanol at elevated pressure and low to intermediate temperature conditions.

## 2. Experimental

The Rapid Compression Machine used in the current study has been described elsewhere [11]. The basic details are provided here for reference. The present RCM is a pneumatically-driven/hydraulically-stopped arrangement, which provides for compression times on the order of 30 ms. The states in the reaction chamber when the piston reaches Top Dead Center (TDC) are referred to as the compressed conditions. The initial temperature, initial pressure, and compression ratio can be varied to vary the compressed temperature ($T_C$) and compressed pressure ($P_C$) independently.

Fuel/oxidizer premixtures were made in a 17 L mixing tank, equipped with heaters and a magnetic stirring apparatus. The reaction chamber of the RCM was also heated, allowing the entire system to reach temperatures up to 140 °C. This allows fuels with rather low vapor pressure to be studied in the RCM. The fuels used in this study were *n*-butanol (anhydrous, 99.9%), *iso*-butanol (99.5%), *sec*-butanol (99.5%), and *tert*-butanol (99.7%), while $O_2$ (99.8%) and $N_2$ (99.998%) were used to create the oxidizer. *n*-, *iso*-, and *sec*-butanol are liquids at room temperature and have relatively low vapor pressure, so they were massed gravimetrically in a syringe to within 0.01 g of the specified value. *tert*-Butanol is a solid at room temperature and was first melted in a glass container before being massed in the same manner as the rest of the fuels. Proportions of the gases in the mixture were determined manometrically and added at room temperature. The saturation vapor dependence of the fuels was taken from the *Chemical Properties Handbook* by Yaws [12]. The preheat temperature of the mixing tank was set above the saturation temperature of the fuels to ensure their complete vaporization.

One of the most important considerations is to ensure that the fuel and oxidizer are uniformly mixed to ensure homogeneous conditions for all the experiments. This was accomplished by heating the system over the course of approximately two hours, while simultaneously applying the magnetic stirrer. Tests with Gas Chromatography/Mass Spectrometry were also conducted to ensure that there was no thermal decomposition of the fuel in the mixing tank and the expected mixture was present in the mixing tank for the entire duration of the experiments.

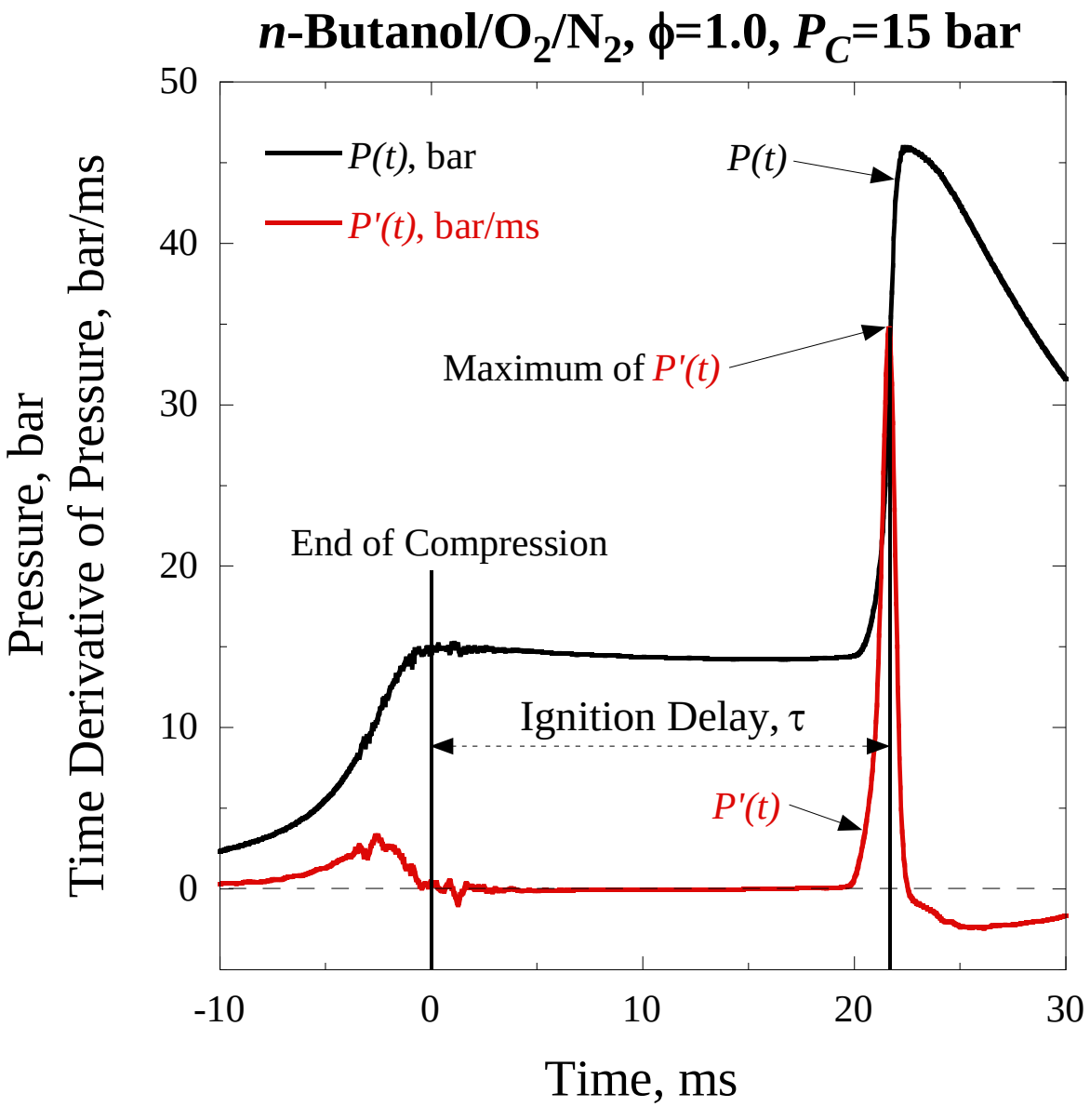

**Figure 1. Definition of ignition delay used in this study. $P'(t)$ is the time derivative of the pressure.**

Experiments were carried out at the same pressure and equivalence ratio condition for all four isomers of butanol. The compressed pressure ($P_C$) condition was chosen to provide data at engine relevant conditions, in a range that has not been covered previously. All experiments were carried out at $P_C$=15 bar, for $\phi$=1.0 mixture in nitrogen-oxygen air. The corresponding reactant mole fractions were: $X_{fuel} = 0.0338$, $X_{O2} = 0.2030$, and $X_{N2} = 0.7632$. The compressed temperature ($T_C$) conditions were similar for all the fuels, ranging from 725 K to 870 K.

The end of compression, when the piston reached TDC, was identified by the maximum of the pressure trace ($P(t)$) prior to the ignition point. The local maximum of the derivative of the pressure trace with respect to time ($P'(t)$), in the time after

TDC, was defined as the point of ignition. The ignition delay was the time difference between the point of ignition and the end of compression. Figure 1 illustrates the definition of ignition delay ($\tau$) used in this study.

Each compressed pressure and temperature condition was repeated at least six times to ensure reproducibility. The mean and standard deviation of the ignition delay for all concurrent runs were calculated; as an indication of reproducibility, one standard deviation of the ignition delays was less than 10% of the mean in all cases. Representative experimental pressure traces for simulations and plotting were chosen as the run whose ignition delay was closest to the mean. Furthermore, each new mixture preparation was checked against previously tested conditions to ensure consistency.

Two types of simulations were performed using CHEMKIN-PRO [13]. The first was a constant volume, adiabatic simulation, whose initial conditions were set to the compressed conditions in the reaction chamber. The second type was a variable volume simulation, where the volume of the simulated reaction chamber was a controlled function of time, so that the simulated pressure trace matched the experimental trace both during and after compression. Heat loss during and after compression were modeled empirically to fit the experimental pressure trace of the corresponding non-reactive pressure trace, as described in Ref. [11]. A non-reactive pressure trace was obtained by replacing oxygen with nitrogen in the mixture. This replacement maintained a similar mixture specific heat ratio, while eliminating oxidation reactions that can cause major heat release.

Temperature at TDC was used as the reference temperature for reporting ignition delay data and was called the compressed temperature ($T_C$). The temperature was calculated using the variable volume simulations. The kinetic mechanisms used in this study were taken from the work by Moss et al. [7], Grana et al. [8], and Van Geem et al. [9]. To ensure no significant chemical heat release was contributing to the determination of the temperature at TDC, calculations were performed and compared with and without reaction steps for each kinetic mechanism; the temperature profile during the compression stroke was the same whether or not reactions were included. This approach has been validated in Refs. [11,14].

## 3. Discussion

Figures 2(a)-2(d) show the experimental pressure traces from the RCM for the four isomers of butanol, with the compressed temperature for each run labeled on the figures. The non-reactive case, described previously, is a run with oxygen in the mixture replaced by nitrogen to suppress oxidation reactions but maintain a similar specific heat ratio. These figures show one of the primary advantages of the RCM, namely, the ability to maintain nearly constant compressed pressure over a range of compressed temperatures. Each of the fuels has monotonically decreasing ignition delay with increasing temperature, indicating there is no NTC region present in this temperature and pressure range. In addition, there is clearly no evidence of two-stage ignition for any of these fuels under the conditions investigated.

Furthermore, for *sec*-, *tert*-, and *iso*-butanol, the non-reactive pressure trace closely matches the reactive cases, up until the point of hot ignition. This indicates there is little to no pre-ignition heat release. By contrast, there is a clear deviation of the non-reactive trace from the reactive traces in the case of *n*-butanol, indicating some chemical heat release prior to hot ignition.

Figure 3 shows an Arrhenius plot of the ignition delays of the four isomers of butanol. The vertical error bars represent two standard deviations of the ignition delay, calculated from all the runs at that condition; the dashed lines are least squares fits to the data.

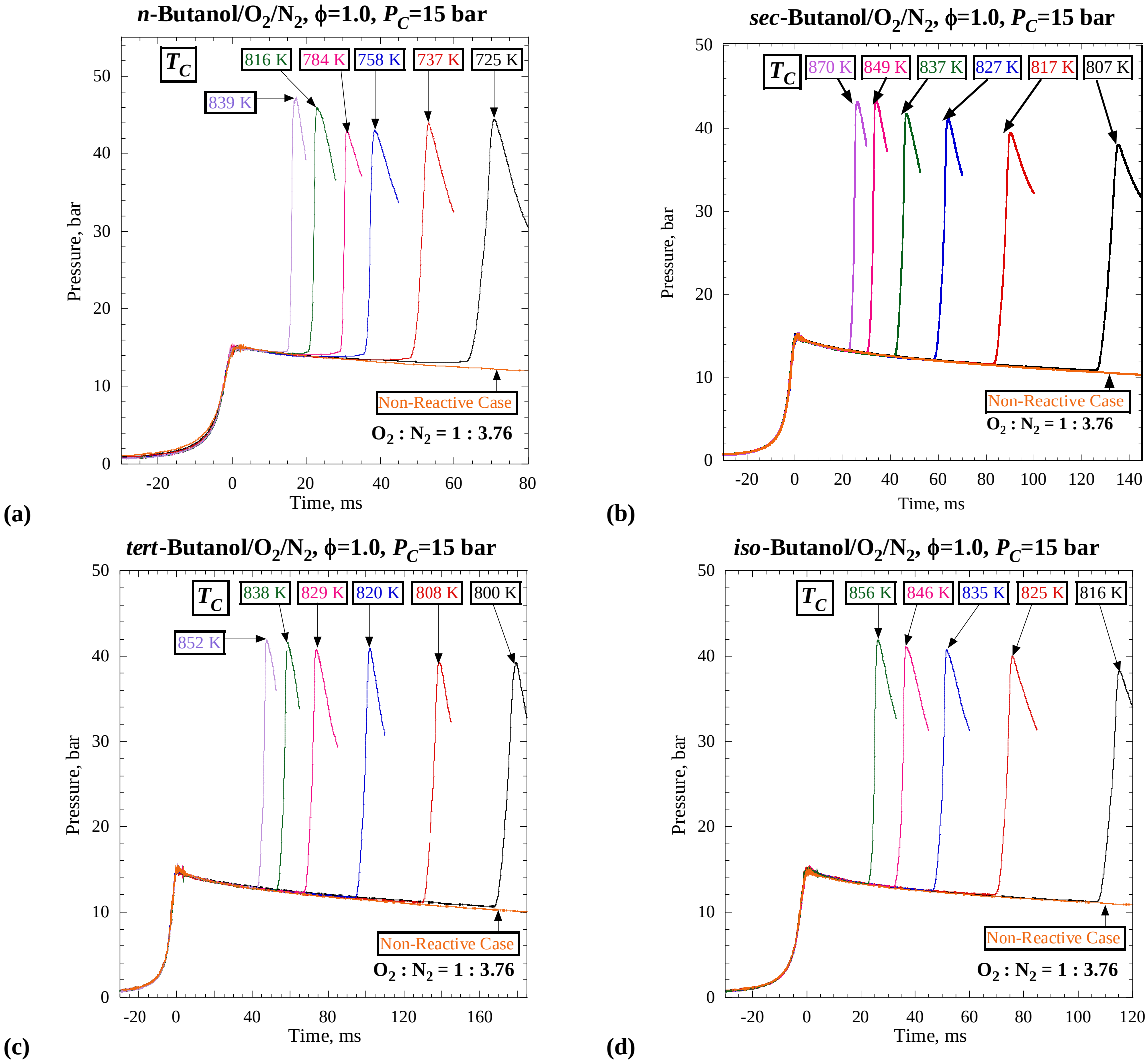

**Figure 2. Experimental pressure traces in the RCM for the four isomers of butanol. Note the absence of NTC and two-stage ignition on these plots.**

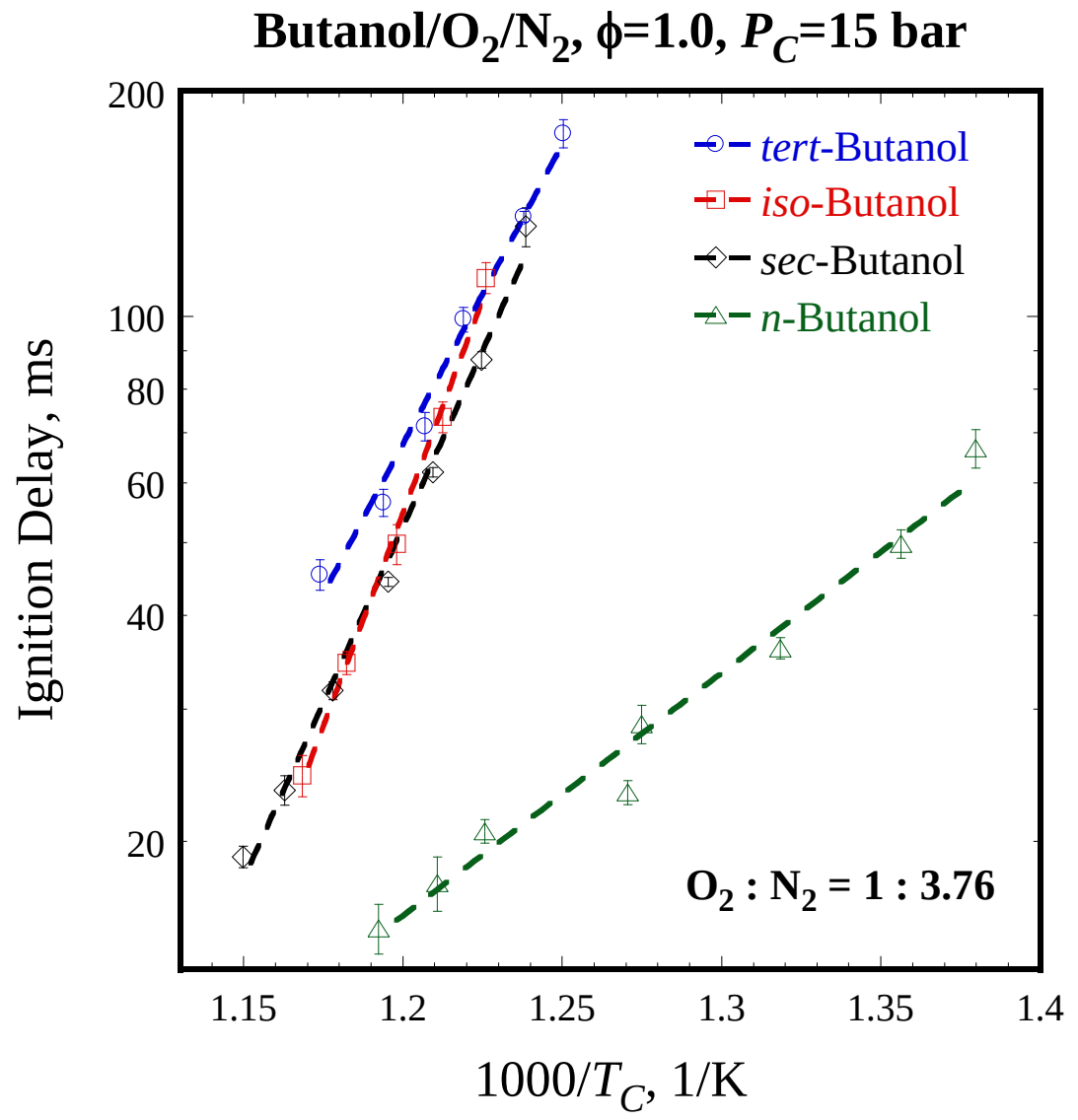

**Figure 3. Arrhenius plot of the ignition delays of the four isomers of butanol.**

Figure 3 demonstrates quite clearly the differences in reactivity between the fuels. *n*-Butanol is clearly the most reactive, followed by *sec*- and *iso*-butanol, which have very similar reactivities, and *tert*-butanol. The extremes of this list agree with the results found previously in studies such as that by Moss et al. [7] and Veloo and Egolfopolous [10] – *n*-butanol is the most reactive of the butanol isomers, and *tert*-butanol is the least reactive. The two intermediate isomers show significant overlap in their ignition delays in this temperature range, making a distinct determination of greater reactivity more ambiguous. This is in contrast to the studies by Moss et al. [7] and Veloo and Egolfopolous [10] who found distinct differences in the reactivities for *iso*- and *sec*-butanol. Specifically, they found that *sec*-butanol is more reactive than *iso*-butanol in the temperature range they were studying. In the current experiment, it appears that *iso*-butanol does not become less reactive than *sec*-butanol until approximately 830 K, and *sec*-butanol continues to become relatively less reactive as temperature increases. However, they are really so close that it is difficult to draw distinct conclusions.

The activation energies of *sec*-, and *iso*-butanol are similar in this temperature range, but the activation energy of *tert*-butanol appears to be slightly lower than the other two. This causes an apparent crossover of the ignition delay between 800 K and 820 K. In this range, as temperature continues to decrease, *tert*-butanol apparently becomes more reactive than first *iso*- and then *sec*-butanol. Future data sets are planned to extend the data to lower temperatures to systematically investigate this feature.

Figures 4(a)-4(d) show the ignition delays of the four isomers of butanol compared against simulations using three mechanisms available in the literature. Data points represent the current experiments, with vertical error bars equal to twice the standard deviation of the ignition delays, as described previously. The dashed lines are least squares fits to the data, the solid lines are constant volume, adiabatic simulations, and when included, the dotted lines are "volume as a function of time", or variable volume simulations.

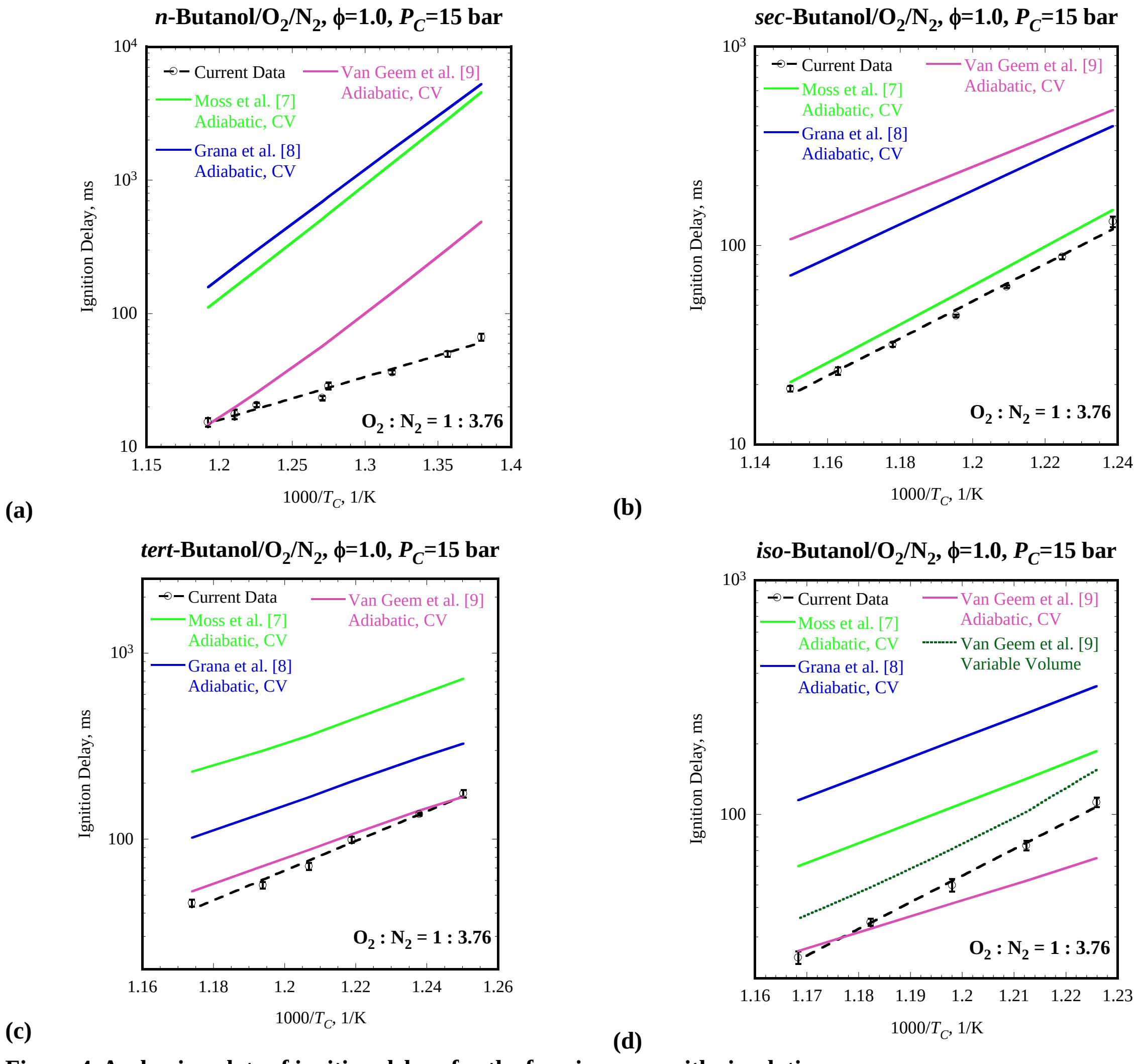

**Figure 4. Arrhenius plots of ignition delays for the four isomers, with simulations.**

Using constant volume, adiabatic simulations, the mechanisms from Moss et al. [7] and Grana et al.[8] over-predict the ignition delay for all four isomers of butanol. This is probably because neither mechanism includes low-temperature chemistry of the butanols. However, it is interesting to note that simulations using the mechanism by Moss et al. [7] predict the ignition delay of *sec*-butanol closely, and reproduce the apparent overall activation energy quite well.

Using constant volume, adiabatic simulations, the mechanism from Van Geem et al. [9] over-predicts the ignition delay for *n*-, *sec*-, and *tert*-butanol, but under-predicts the ignition delay for *iso*-butanol. The simulations are quite close to the experimental values over the whole experimental range for *iso*- and *tert*-butanol, and for the higher temperatures of the experimental range of *n*-butanol. Variable volume simulations were computed for *iso*-butanol, since the experimental values of the ignition delay were under-predicted by the mechanism from Van Geem et al. [9]. Although the variable volume simulations over-predict the ignition delay, they improve the prediction of the apparent overall activation energy. It is also interesting to note that the order of reactivity of the three mechanisms differs in this temperature and pressure range.

## 4. Conclusions

The autoignition delay of the four isomers of butanol has been measured in a Rapid Compression Machine, at a compressed pressure of 15 bar and compressed temperatures ranging from 725 K to 870 K. The stoichiometric condition, in nitrogen/oxygen air, was studied for all four fuels. The reactivity of the isomers of butanol in this temperature and pressure range was found to be: $n$-butanol $>$ $sec$-butanol $\sim$ $iso$-butanol $>$ $tert$-butanol, but this ranking appears to be a function of temperature. Simulations using three mechanisms available in the literature generally over-predicted the ignition delays.

### Acknowledgments

This material is based upon work supported as part of the Combustion Energy Frontier Research Center, an Energy Frontier Research Center funded by the U.S. Department of Energy, Office of Science, Office of Basic Energy Sciences, under Award Number DE-SC0001198.